\documentclass[12pt,preprint]{aastex}

\newcommand{\st}{\scriptscriptstyle}
\newcommand{\EE}[2]{{#1$\times$$10^{#2}$\/}}
\newcommand{\e}[1]{{$10^{#1}$\/}}  
\newcommand{\etal}{et al.}
\newcommand{\msun}{{\rm M}_\odot}
\newcommand{\pc}{\,{\rm pc}}

\newcommand{\ergs}{\,{\rm ergs}}
\newcommand{\K}{\,{\rm K}}

\newcommand{\mswept}{$M_{\rm swept}(\msun)$}
\newcommand{\rsh}{$R_{s}$}
\newcommand{\vsh}{$V_{s}$}

\begin{document}

\title{Supernova Blastwaves in Low-density Hot Media: a Mechanism for
Spatially Distributed Heating}
\author{Shikui Tang and Q. Daniel Wang}
\affil{Department of Astronomy, University of Massachusetts,
  Amherst, MA 01003\\
  tangsk@astro.umass.edu and wqd@astro.umass.edu}

\begin{abstract}
Most supernovae are expected to explode in low-density hot media, 
particularly in galactic bulges and elliptical galaxies. 
The remnants of such supernovae, though difficult to detect individually, can be
profoundly important in heating the media on large scales. We characterize the
evolution of this kind of supernova remnants, based on analytical 
approximations and hydrodynamic simulations. We generalize the standard
Sedov solution to account for both temperature and density effects of
the ambient media. Although cooling can be neglected, the expansion of 
such a remnant deviates quickly
from the standard Sedov solution and asymptotically approaches the ambient sound
speed as the swept-up thermal energy becomes important. The relatively steady
and fast expansion of the remnants over large volumes provides an ideal
mechanism for spatially distributed heating, which may help to alleviate
the over-cooling problem of hot gas
in groups and clusters of galaxies as well as in galaxies themselves. The 
simulations were performed with the FLASH code.

\end{abstract}

\section{Introduction}

Supernovae (SNe) are a major source of the mechanical energy input in galaxies
and possibly in the intergalactic medium (IGM). On average, an SN releases about
$10^{51}\ergs$ of kinetic energy carried by the ejecta which drive a blastwave
into the ambient medium. How far  this blastwave goes and how fast
the energy is dissipated depend sensitively on the density and temperature of
the medium. Core-collapsed SNe represent the end of
massive stars, the bulk (if not all) of which form in OB associations. 
The energy release from such an association, highly correlated in space and 
time, has a great impact on the surrounding medium. Initially, the energy 
release from the OB association is primarily 
in form of intense ionizing radiation from very massive stars, which 
tends to homogenize the surrounding medium \citep{McKee1984}.
After a couple of $10^6$ years, fast stellar winds start to play a major
role in heating and shaping the medium, creating a low-density hot bubble 
\citep{Weaver1977}, probably even before the explosion of the 
first SN in the association.
Later, after about $5 \times 10^6$ years of the star formation (if 
more-or-less coeval), 
core-collapsed SNe become the dominant source of the mechanical 
energy input into the already hot surrounding \citep{Monaco2004}. 
This combination
of the concerted feedbacks, lasting for $\sim 5 \times 10^7$ 
years --- the lifetime of an 8$M_\odot$ star, leads to the formation 
of a so-called superbubble
of low-density hot gas enclosed by a supershell of swept-up cool gas 
(e.g., \citealt{Mac1988}). 
The expansion of such a superbubble is expected
to be substantially faster than typical OB association 
internal velocities of a few ${\rm~km~s^{-1}}$. Therefore, a majority
of core-collapsed SNe ($\sim 90\%$) should occur inside their parent 
superbubbles (e.g., \citealt{Higdon1998}; 
\citealt{Parizot2004} and references therein). SNe from lower mass stars
(e.g., Type Ia; e.g., \citealt{McMillan1996}), particularly important 
in early-type galaxies and possibly in galactic bulges and halos, are also
expected to occur mostly in low-density hot media. The interstellar
remnants of such SNe are typically too faint to be well observed individually.
But in terms of both heating and shaping the global ISM/IGM,  
such ``missing'' remnants are probably more important 
than those commonly-known and well-studied supernova remnants
(SNRs). Though spectacular looking, they are atypical products of SNe 
(e.g., from run-away massive stars), which happened to be in relatively 
dense media.

Surprisingly, there has been little work toward the understanding of the 
SNR evolution in hot media with relatively high pressure. 
Almost all the existing studies assume an ambient medium with 
both low temperature and low pressure
(e.g., $T\lesssim10^4\K$, $nT\lesssim10^{4}{\rm~cm^{-3}K}$; 
\citealp{Chevalier1974, Cioffi1988, Shelton1999}) 
or relative high in temperature but still low in pressure
(e.g, $T\sim 10^{6-7}\K$, $nT \sim 10^{3-4}{\rm~cm^{-3}K}$; \citealt{Dorfi1996}). 
If both the ambient gas temperature and pressure are low 
(e.g., $T \lesssim 10^{4}{\rm~K}$, $nT \lesssim 10^4{\rm~cm^{-3}K}$ ), 
much of the SNR evolution may then be described by the self-similar 
Sedov solution \citep{Sedov1959}, which assumes that 
both the thermal energy and radiative cooling of swept-up gas are negligible. 
The ambient gas with both high temperature and high pressure makes two differences: 
(1) the shock is weak due to the high sound speed and 
(2) thermal energy of the swept-up gas is not negligible. 
Sedov (1959, P238-251) made a linear approximation to the pressure effect, 
but did not account for the high temperature effect \citep{Gaffet1978}. 
\citet{Dorfi1996} briefly mentioned the behavior of 
an SNR shock asymptotically reaching the sound speed, 
but studied primarily the cosmic ray acceleration.
\citet{Mc} considered the SNR evolution in a cloudy medium, including
thermal evaporation. Although the inter-cloud medium in this case 
is hot ($T\sim 10^6\K$), the average ambient pressure is still low 
($nT\sim 10^3{\rm~cm^{-3}K}$). 
The characteristic temperature and pressure inside galactic bulges 
and elliptical galaxies are about 2$\sim$3 orders of magnitude higher than 
the values used in these SNR studies 
(\citealp{Spergel1992,Morris1996} and references therein).
\citet{Shigeyama1997} simulated an SNR in a hot medium with relative high pressure 
($n\! =\! 0.1 {\rm cm^{-3}}$ and $T\!\approx\!10^7\K$), but focused only on the ejecta. 
In short, we are not aware of any simple description or simulation of
the SNR evolution in the hot media typically for galactic bulge and elliptical galaxies.

We have closely examined the SNR evolution in the relatively low-density 
and high pressure hot media.
The evolution of such an SNR is significantly different from
those in a relatively dense and cold environment. While the low density means 
that the SN energy loss rate is slow, the high pressure makes the swept-up
thermal energy dynamically important. Furthermore, the expansion
speed of the blastwave is always above the sound speed of the hot ambient 
medium with the value of a few hundred $\rm km\,s^{-1}$.
Therefore, the SNR can expand to a large volume and distribute its energy in
a rather uniform fashion. In the following, we demonstrate these effects, based chiefly on
1-D high-resolution hydrodynamic simulations.

\section{Simulation Setup}

Our simulations use the newly released FLASH code (version 2.4),
which allows for modular, adaptive-mesh, and parallel simulations and
solves the hydrodynamic equation explicitly (\citealt{Fryxell2000} and references therein)
To accurately capture a shock front, we use a uniformly spaced grid with a
spatial resolution of about 0.01\,pc.
We assume that the SN explosion is spherically symmetric and the ambient medium
is uniformly distributed with identical temperature ($T_0$) and 
density ($\rho_0$=$n_0 m_p$). The number density of particles is $n = n_0/\mu$
where $\mu \approx 0.6$ for fully ionized gas with the solar abundance. An SN 
energy $E_{sn}$ ($= 10^{51} \ergs$) is
deposited inside a small radius $r_{init}$. 
The ambient gas is assumed to be mostly ionized. The cooling is neglected in the simulations (further
discussion in \S 4). Table \ref{T:parameters} lists the setup parameters 
($T_0$, $n_0$ and $r_{init}$) 
of a few representative cases, which are 
characteristic of the Galactic bulge (A), giant elliptical galaxies (B),
and rich clusters of galaxies (C). The table also lists various 
inferred parameters to be discussed in the following sections.

We have experimented with various energy deposition schemes
\citep{Cioffi1988,Chevalier1974}. If the explosion energy is initially
deposited as heat uniformly in a small volume, a Sedov solution can be quickly 
reached. On the other hand, depositing the energy in the form of kinetic energy leads to 
many small-scale structures due to various internal shocks especially inside the ejecta. 
With or without ejecta (assumed to have a total mass of 1.4 $M_\odot$) 
do affect the evolution before the swept-up mass becomes dominant compared with the ejecta. 
However, a specific choice of the initial conditions does not significantly 
affect our conclusion on the overall structure and evolution of the SNRs.

We output simulation results every 100 years. 
The radius of the outer shock front (\rsh) corresponds to 
the position where the gradients of the
pressure and velocity are the largest. 
The shock front velocity (\vsh) is estimated from
the \rsh\ difference between two consecutive steps. Therefore, the accuracy of
the local \rsh\ and \vsh\  estimates is limited by the finite resolution in both
time and spatial step sizes. An adaptive smooth of the \vsh\
evolution is performed to reduce the step-by-step fluctuation. The
uncertainties in these calculations in individual steps do not affect the
actual \textit{evolution} of these parameters, which are determined by the 
internal hydrodynamic solutions in the simulations.

\section{Results: Outer Blastwave Evolution}

\begin{figure}
\begin{center}
\plottwo{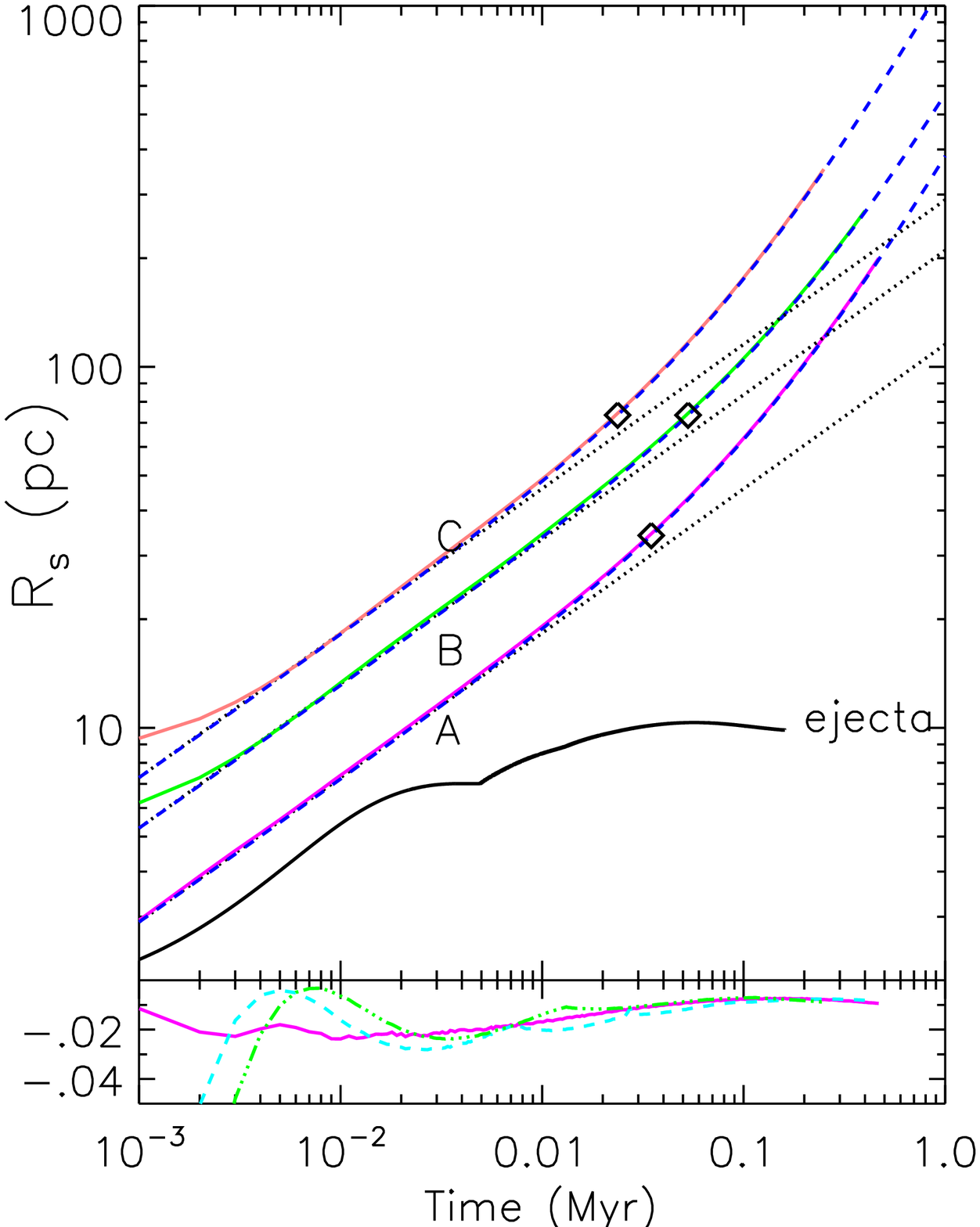}{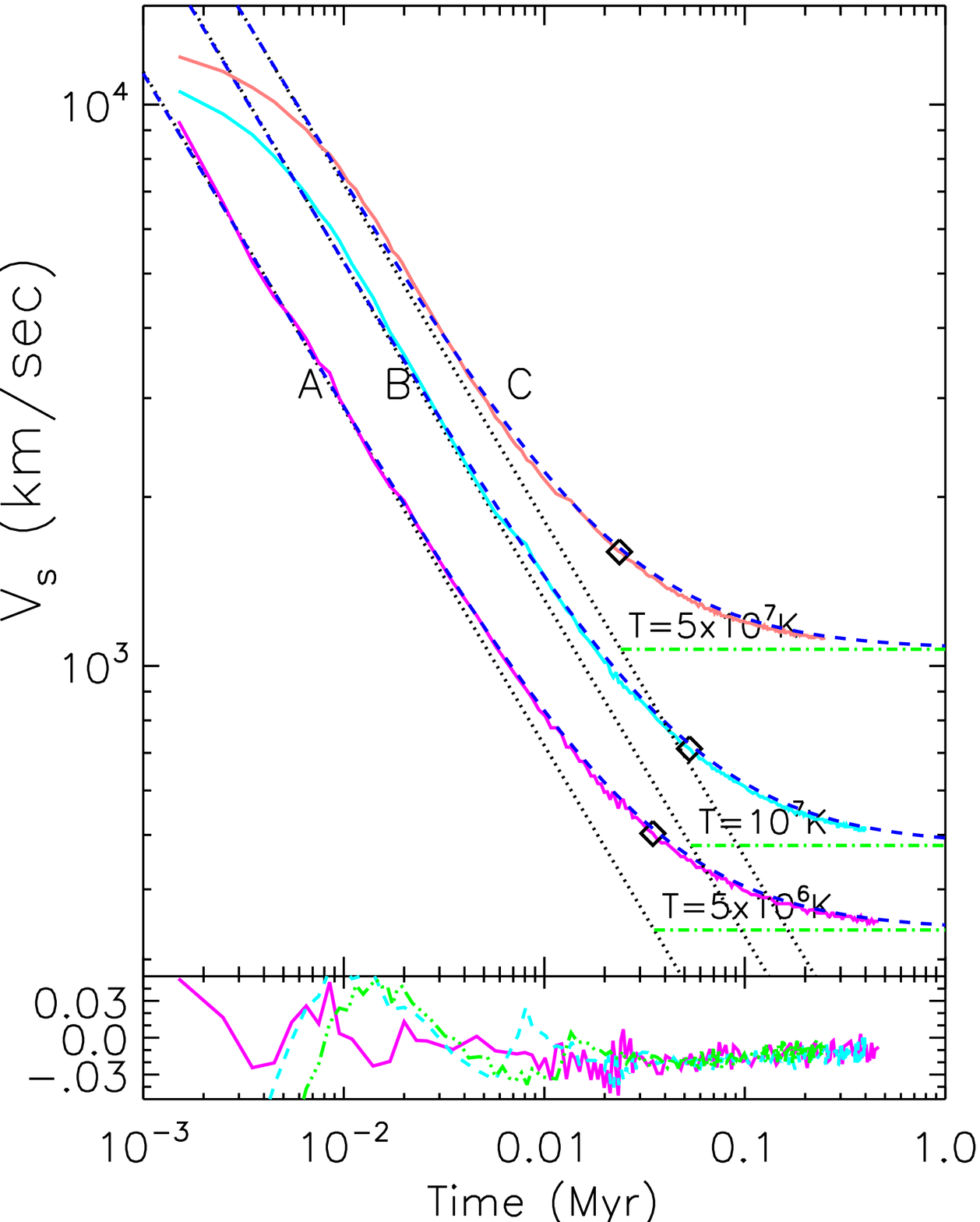}
\caption{\label{Fig:cmp_rsh_vsh}   The evolution of \rsh\ (left-hand panel) and
\vsh\ (right-hand panel) for three cases. The colored (light) 
solid curves, marked with Case A, B, and C (Table 1), are the simulation results; dotted straight lines are the corresponding analytic Sedov solutions. The heavy solid curve in the right panel represents the evolution of the contact discontinuity.  The
diamond marks the values of \rsh\ and \vsh\ at time $t_c$. The horizontal dash-dotted
lines in the right-hand panel denote the corresponding sound velocities of ambient media. The bottom of each panel shows the relative error of the simulated results to the generalized analytic approximations (Eqs. \ref{eq:vsh} and ~\ref{eq:rsh}) for Case A (solid curve), B (dashed curve), and C (dash-dot-dotted
curve).}
\end{center}
\end{figure}

Figure \ref{Fig:cmp_rsh_vsh} shows the evolution of \rsh\ and \vsh\ for the three 
cases listed in Table \ref{T:parameters}. 
In all cases, the initial free-expansion stage is short and ends 
when the swept-up mass roughly equals to that of the ejecta. 
The expansion then more-or-less follows the self-similar Sedov solution:
\begin{equation}
V_{Sedov} = \frac{2}{5}\xi\left(\frac{E_{sn}}{\rho_0 t^3}\right)^{1/5}, 
\label{eq:vs_sedov}
\end{equation}
where $\xi$=1.14 for gas with $\gamma$=5/3. 
But, the Sedov phase does not last long if the ambient medium has a high temperature. 
The evolution gradually deviates from the Sedov solution, as the blastwave expansion 
asymptotically approaches the sound speed of the ambient medium.
Based on these asymptotical behaviors (neglecting the brief free-expansion 
phase), we find that the following 
expression gives a simple generalization of the Sedov solution, 
approximately characterizing the blastwave evolution in a low-density hot medium 
(to an accuracy of $\lesssim 3\%$; see the bottom panels of Figure \ref{Fig:cmp_rsh_vsh}):
\begin{equation}
V_s = c_s \left( \frac{t_c}{t}+1\right)^{3/5}
\label{eq:vsh}
\end{equation}
where $c_s$ is the sound speed of the ambient medium, and $t_c$ is 
a characteristic time which can be obtained from the Sedov solution 
by equaling $V_{Sedov}$ and  $V_s$ when $t \ll t_c$: 
\begin{equation}
t_c = \left[ \left(\frac{2}{5}\xi\right)^5 \frac{E_{sn}}{\rho_0 c_s^5} \right]^{1/3}.
\label{eq:tc}
\end{equation}

From Eq. (\ref{eq:vsh}), one can easily estimate the Mach number $M = V_s/c_s$ of the
blastwave as function of time, e.g., $M\approx 1.5$ at $t=t_c$ and a 
strong shock (i.e., $M>2$) for $t < 0.46t_c$, or about a few $10^4$ years. 
For ease of reference, Table~\ref{T:parameters} also lists
the time and the swept-up ambient mass when $M=2$. Note that Eq. (\ref{eq:vs_sedov}) 
is not valid even before $t=t_c$. Subsequently, the blastwave
expands with a low Mach number and will eventually be dissipated by 
radiative cooling and turbulent motion, which cannot be accounted 
for here in 1-D simulations, however. 
But the expansion speed would not fall below the sound speed. 

By integrating Eq.~\ref{eq:vsh}, we further derive the blastwave radius,
\begin{eqnarray}
R_{s}(t) &=&\int_{0}^{t} c_s \left( \frac{t_c}{t'}+1\right)^{3/5} dt'  \nonumber \\
&=&  \frac{5}{2}c_s{t_c}
\left(\frac{t}{t_c}\right)^{2/5}
F\left(-\frac{3}{5},\frac{2}{5};\frac{7}{5};-\frac{t}{t_c}\right),
\label{eq:rsh}
\end{eqnarray}
where $F$ is the generalized hyper-geometric function.

The characteristic parameter $t_c$ has a clear physical meaning.
Within the radius $R_c \equiv R_{s}(t_c) \approx 2.89 c_s t_c$,
the ambient thermal energy swept up by the blastwave is
\begin{equation}
\frac{4}{3}\pi R_c^3 \frac{n_0 k T_0}{\mu (\gamma-1)}=
\frac{0.247\pi \xi^5}{\gamma(\gamma-1)}
E_{sn}
\simeq  1.8 E_{sn},
\label{eq:phy_tc}
\end{equation}
where $\mu$=0.6.
Therefore, $t_c$ characterizes the time when the swept-up thermal energy is
about twice the explosion energy. Note that for a given $E_{sn}$, 
$R_c$ depends only on the ambient pressure, 
but $t_c$ depends on both the density and the temperature. 

Eqs. (\ref{eq:vsh}) and (\ref{eq:rsh}) generalize the Sedov solution,
by accounting for both temperature and density effects of the ambient medium.
The temperature effect is reflected in the explicit dependence of the 
SNR evolution on $c_s$. When $c_s\rightarrow 0$ (thus $t_c \rightarrow \infty$), Eqs. (\ref{eq:vsh}) 
and (\ref{eq:rsh}) become the Sedov solution. In general, 
$R_s$ is determined by $M$, which depends only on $t/t_c$ 
(Eq. \ref{eq:vsh}), and $R_c$.
In other words, $M$ is uniquely determined by $R_s$ for a given $R_c$.
Eqs. (\ref{eq:vsh}) and (\ref{eq:rsh}) show that both \vsh\ and
\rsh\ could have substantially larger values than what are predicted by
the Sedov solution, while $M$ is generally low. Therefore, 
the SNR blastwave heating is locally gentle and
is over a large volume in the low-density hot medium. 

\section{Results: Interior Structure of the Supernova Remnants}

\begin{figure}
\begin{center}
\plotone{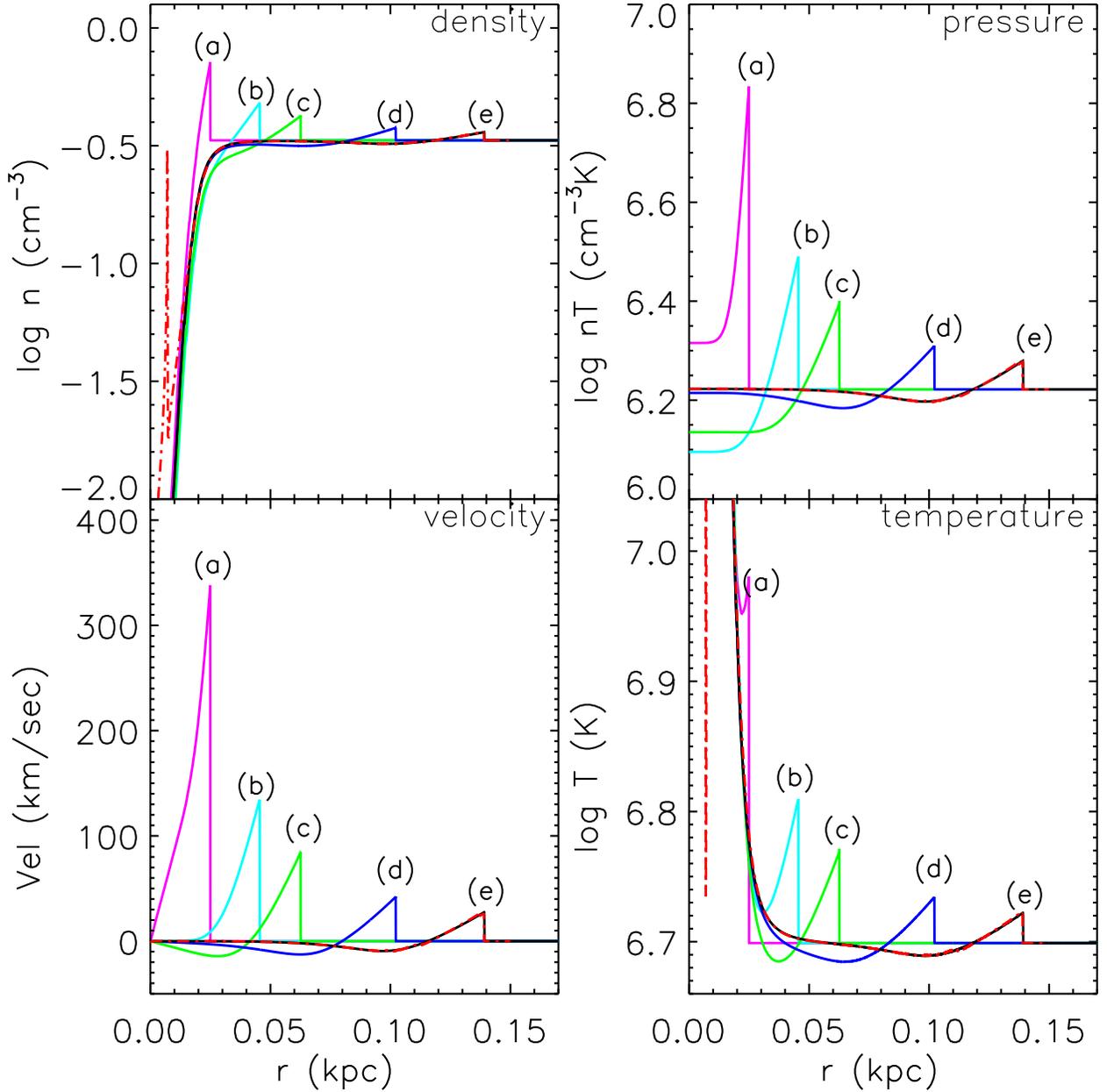}
\caption {\label{Fig:snrs_structure} 
Solid curves show the structures of Case A with energy deposited as pure thermal energy at 
(a) 0.018/1.87, (b) 0.058/1.33, (c) 0.098/1.20, (d) 0.20/1.10,
and (e) 0.30/1.07 Myr/Mach number. For comparison, the dash-dotted (red) line illustrates the
result at the same time as (e) from a simulation in which $E_{sn}$ is deposited in form of kinetic energy.} 
\end{center}
\end{figure}

\begin{figure}
\begin{center}
\plotone{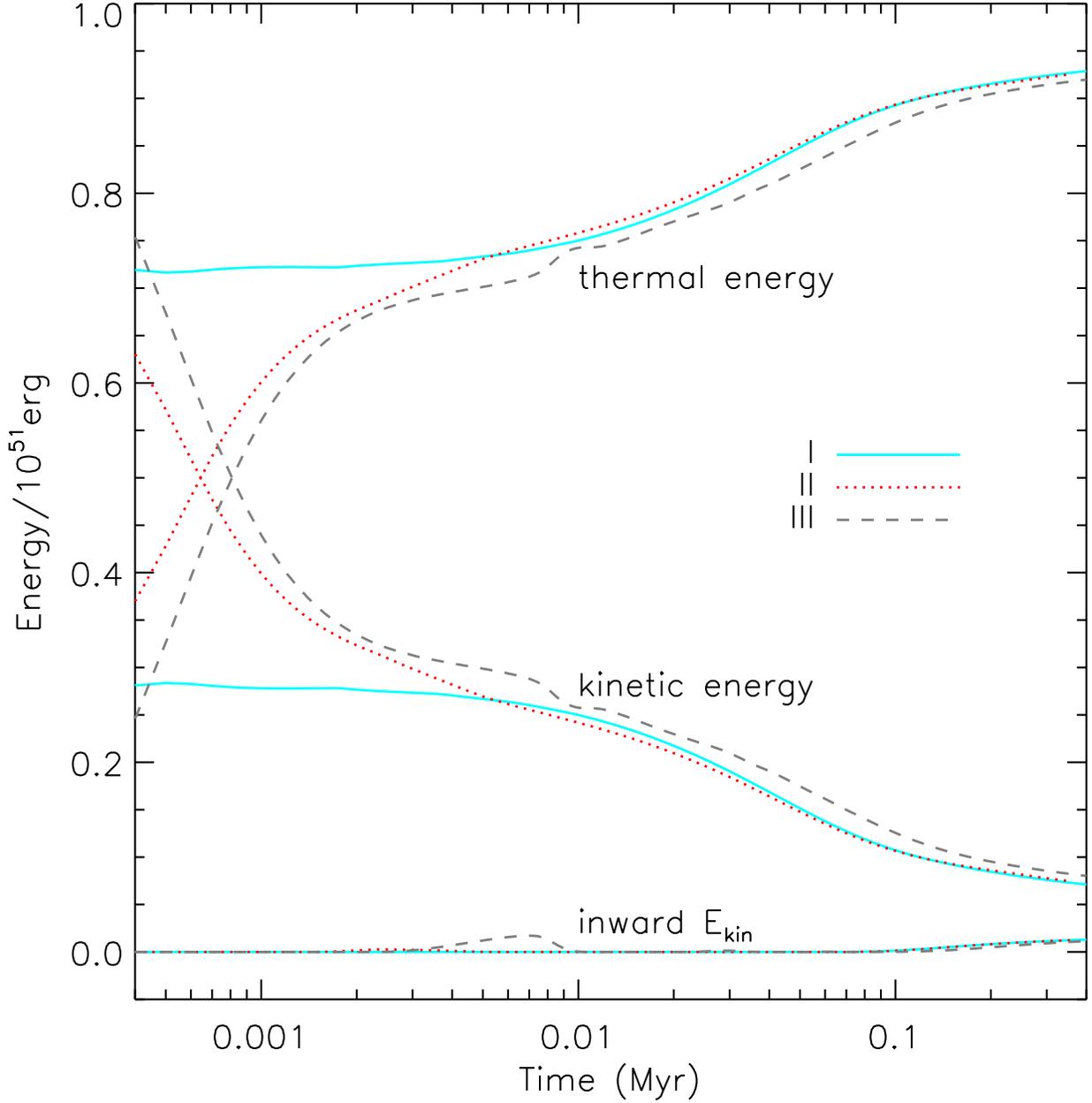}
\caption { \label{Fig:ener_evolution} The evolution of thermal and
kinetic energies of Case A SNR: I --- $E_{sn}$ as pure
thermal energy without ejecta; II --- $E_{sn}$ as pure thermal energy
with 1.4 $M_\odot$ ejecta; and III --- $E_{sn}$ as kinetic energy with 1.4 $M_\odot$ ejecta.}
\end{center}
\end{figure}

Figure~\ref{Fig:snrs_structure} illustrates the radial structure of our
simulated SNRs (Case A as an example). At the early stage (before time a) 
the SNR maintains a strong shock and the post-shock gas moves forward at 
a speed greater than the sound speed of the ambient gas. A low pressure region
gradually develops inside the SNR due to the adiabatic expansion, as shown in
the pressure panel. Remarkably, both density and temperatures can fall below 
the ambient values. Therefore, the cooling rate, hence the X-ray emission, of this 
region can be very low. The pressure at the SNR center reaches the 
minimum at time b. Later, the post-shock gas starts to flow back and the
central pressure increases, gradually approaching the value of the ambient 
medium (time c-e). But, the under-pressure region behind the widening blastwave front remains. 
Figure~\ref{Fig:ener_evolution} shows the conversion of the explosion energy to the
thermal and kinetic energy of the swept-up medium as a function of time, which
depends weakly on the specific
initial conditions except for the first several $10^3$ years (\S 2). 
In addition to the explosion energy, the blastwave also redistributes the 
thermal energy of the swept-up ambient medium.

While the energies are  redistributed widely, the ejecta are
confined within relatively small regions. Figure~\ref{Fig:cmp_rsh_vsh}
(left-hand panel) includes the evolution of the contact discontinuity 
between the ejecta and the swept-up ambient medium of Case A with $1.4\,\msun$ ejecta. The 
discontinuity reaches a maximum of only $\sim 10\pc$, consistent with the result 
of \citet{Shigeyama1997}. Therefore, the metal enrichment of an SNR 
may still be localized even in a low density medium, 
unless an interaction with another SNR or a global flow such as a galactic wind redistributes the ejecta.

We have neglected the cooling effect. Using the collisional ionization 
equilibrium (CIE) cooling rate, we estimate the total radiative energy loss 
to be less than 5\% up to 0.2 Myr.
The undisturbed ambient gas itself radiates even more efficiently than 
the swept-up gas by SN blastwave.
The potential deviation from the assumed CIE should 
be small, because the ambient medium considered here is already highly ionized. 
 
\section{Discussions}

The above results show that the evolution of an SNR in a low-density hot medium
has several distinct properties: 1) The blastwave always moves 
at a speed greater than, or comparable to, the sound wave and can thus reach 
a much larger radius than that predicted
by the Sedov solution; 
2) Because the remnant never gets into a
snow-plow radiative phase, the radiative  cooling is typically negligible;
3) The swept-up thermal
energy is important, affecting the evolution of both the blastwave and
the interior structure; 4) 
Because the Mach number of the blastwave is typically small, its heating
is subtle and over a large volume.

The large-scale distributed heating by such SNRs
may have strong implications for solving the energy balance
problems in studying diffuse hot gas. The most
notable problem is perhaps the apparent lack of predicted cooling flows 
in groups and clusters of galaxies \citep{Mathews2003, Peterson2003}.
Various existing proposals for the solution (e.g., AGN heating and thermal
conduction) have only limited success and many uncertainties.
Fundamentally, a distributed
heating mechanism is required to balance the cooling
(e.g. \citealt{RRNB2004}).
Indeed, heating due to sound waves generated by buoyant bubbles
from AGN energy injections has recently been proposed as a solution
\citep*{Ruszkowski2004,RBB2004astroph}. But none of these proposals are likely to
explain the more acute over-cooling problem in individual galaxies, where the
cooling time
scale of hot gas is even shorter. As shown above, blastwaves produced
by SNe, Type Ia in particular, may provide a natural mechanism for the
heating required to balance, or at least alleviate, the cooling of the 
diffuse hot gas around
individual galaxies, possibly even in the IGM \citep{Dom2004}.

Clearly, the above discussion is very much limited by our 1-D simulations
of individual SNRs. We are currently carrying out 3-D simulations to study the
evolution of hot gas in galactic bulges. These simulations
will account for the cooling as well as the spatial inhomogeneity
generated by the global galactic gravity and by the interaction among
multiple SNRs. The spatial distribution and physical properties of the gas
will then be determined self-consistently.

\textit{Acknowledgments.} We thank R. A. Chevalier, J. Slavin, R. Shelton,
and the anonymous referee for useful comments on the work, which is
supported by NASA through grant G03-4111X. 
The software used in this work was in part developed 
by the DOE-supported ASC / Alliance Center for Astrophysical 
Thermonuclear Flashes at the University of Chicago.

\clearpage

\clearpage
 
\clearpage

\clearpage

\begin{deluxetable}{c|ccc}
\tabletypesize{\footnotesize}
\tablewidth{0pt} \tablecolumns{4}
\tablecaption{\label{T:parameters} Simulation Parameters}
\tablehead{Parameter & Case A  & Case B  & Case C}
\startdata
$n_0$ &  0.2     &  0.01   &  0.002   \\  
$T_0 \rm (K)$        & \EE{5}{6} & \e{7}  &\EE{5}{7}\\
$r_{init}$(\,pc)     &  1.0     &  5.0    &  2.0    \\ \hline
$c_s \rm (km/sec)$   & 339    & 479  & 1071   \\
$t_c$ (Myr)          & 0.035    & 0.053  & 0.024   \\
\mswept              & 874  & 437  & 87   \\
$t_{\st M = 2}$(Myr)& 0.016 & 0.024     & 0.011   \\
\enddata
\end{deluxetable}

\end{document}